\documentclass{sigchi}

\CopyrightYear{2016}
\setcopyright{acmlicensed}
\usepackage{balance}       % to better equalize the last page
\usepackage{graphics}      % for EPS, load graphicx instead 
\usepackage[T1]{fontenc}   % for umlauts and other diaeresis
\usepackage{txfonts}
\usepackage{mathptmx}
\usepackage[pdflang={en-US},pdftex]{hyperref}
\usepackage{color}
\usepackage{booktabs}
\usepackage{textcomp}
\usepackage{multirow}

\usepackage{microtype}        % Improved Tracking and Kerning
\usepackage{ccicons}          % Cite your images correctly!
\usepackage{todonotes}

\newcommand{\etal}{\textit{et al}. }

\def\plaintitle{PARQR: Augmenting the Piazza Online Forum to Better Support Degree Seeking Online Masters Students}

\def\emptyauthor{}
\def\plainkeywords{Online Forums; Online Degrees; Recommender Systems; Computer-Assisted Instruction; Distance Learning}

\makeatletter
\def\url@leostyle{%
  \@ifundefined{selectfont}{
    \def\UrlFont{\sf}
  }{
    \def\UrlFont{\small\bf\ttfamily}
  }}
\makeatother
\def\pprw{8.5in}
\def\pprh{11in}

\definecolor{linkColor}{RGB}{6,125,233}
\hypersetup{%
  pdftitle={\plaintitle},
  pdfauthor={\emptyauthor},
  pdfkeywords={\plainkeywords},
  pdfdisplaydoctitle=true, % For Accessibility
  bookmarksnumbered,
  pdfstartview={FitH},
  colorlinks,
  citecolor=black,
  filecolor=black,
  linkcolor=black,
  urlcolor=linkColor,
  breaklinks=true,
  hypertexnames=false
}

\usepackage{array,tabularx}
\newenvironment{equation_parameters}
  {\par\vspace{\abovedisplayskip}\noindent\begin{tabular}{>{$}l<{$} @{${}={}$} l}}
  {\end{tabular}\par\vspace{\belowdisplayskip}}

\begin{document}

\title{\plaintitle}

\numberofauthors{1}
\author{
 \alignauthor{Noah Bilgrien, Roy Finkelberg, Chirag Tailor, India Irish, Girish Murali, Abhishek Mangal, Niklas Gustafsson, Sumedha Raman, Thad Starner, Rosa Arriaga\\
    \affaddr{Georgia Institute of Technology}\\
    \affaddr{Atlanta, GA, USA}\\
    \email{\{nbilgrien, roy, chirag.tailor, indiai, girishmn92, abhishekmangal, egustafsson3, sraman46, thad, arriaga\}@gatech.edu}}\\
}

\maketitle
\begin{abstract}
We introduce PARQR, a tool for online education forums that reduces duplicate posts by 40\% in a degree seeking online masters program at a top university. Instead of performing a standard keyword search, PARQR monitors questions as students compose them and continuously suggests relevant posts. In testing, PARQR correctly recommends a relevant post, if one exists, 73.5\% of the time. We discuss PARQR's design, initial experimental results comparing different semesters with and without PARQR, and interviews we conducted with teaching instructors regarding their experience with PARQR.

\end{abstract}

\category{Applied Computing}{Computer-assisted instruction}{}{}
\category{Applied Computing}{Distance Learning}{}{}
  

\section{Introduction}

As online classes play an increasingly substantial role in education~\cite{MOOCStats}, more effort is being exerted to build tools which facilitate education in online classrooms and understand the behavior of students in these classes. However, the majority of these efforts target students in massively open online courses (MOOCs) rather than those in online degree seeking programs. Online degree seeking programs and MOOCS have drastically different expectations and requirements. Students expect and deserve a higher level of human evaluation, feedback, and discussion from their instructors. The recent growth of these programs raises questions regarding how these students behave and interact, how online platforms facilitate learning, discussion, evaluation, and feedback, and how we can build tools to assist in running classes in degree seeking programs at this scale.

Programs utilize online forums and discussion boards such as Piazza and Reddit to facilitate and centralize discussion in the absence of a traditional classroom. However, these forums begin to face problems as class sizes increase. Students are more likely to miss questions that would have benefited them and ask questions which have already been asked. Instructors become inundated with duplicate questions and have difficulty finding the questions which require the most attention. These problems stem from students and instructors not knowing where they should direct their attention due to the volume of material. In a particular class with 391 students using Piazza, we found that 25.6\% of posts were duplicates at the time of posting. That is, an equivalent question had already been asked and answered on the same forum.

We present the Piazza Automated Related Question Recommender (PARQR, pronounced ``parker'') as a solution to this attention attribution problem. PARQR is a recommendation engine which augments the way students and instructors interact with Piazza forums: 1) given a question that a student is writing, it recommends the most similar existing posts 2) it recommends posts most deserving of a student's attention and 3) it recommends posts requiring attention to instructors. We introduced PARQR to the previously mentioned class in which 25.6\% of the posts were duplicates and the percentage of duplicate posts dropped to 17.8\%, a 40\% reduction whose statistical significance is discussed below.

PARQR has been in use by classes at a major public research university since the Spring 2018 semester, being progressively introduced to more classes and having a higher percentage of users in each subsequent class. As of Spring 2019, PARQR has 1000 users across 8 different classes,  approximately $12\%$ of the roughly 8600 currently enrolled in the program.

\begin{figure}[!htb]
\includegraphics[width=0.5\textwidth]{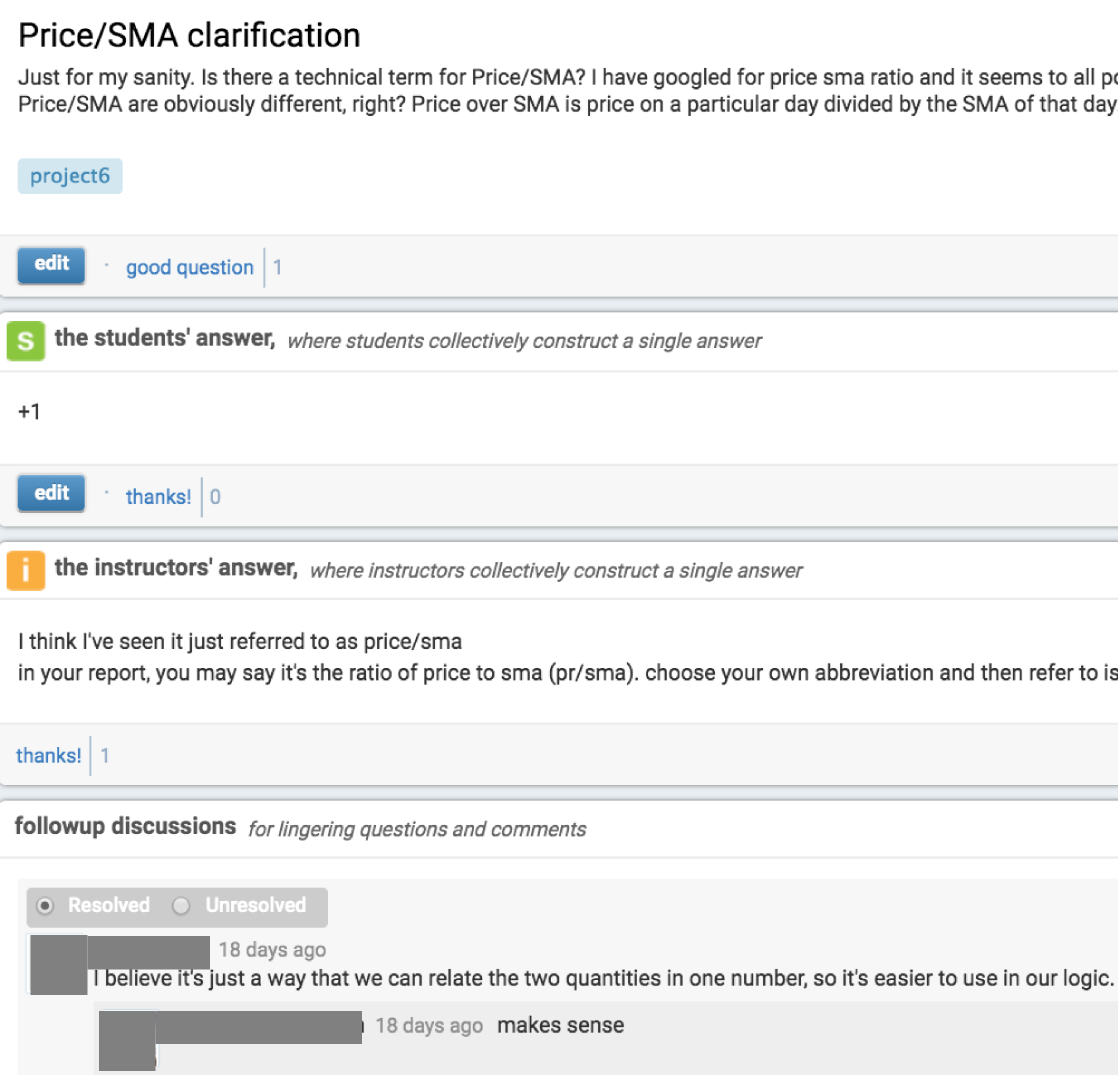}
\caption{\label{fig:piazza-post} A post answered by a student and an instructor along with a followup discussion containing two contributions.}
\vspace{-10pt}
\end{figure}

To better understand the interactions between students, PARQR, and Piazza, we use the following operational definitions. A post is a question created and submitted by a student. Student and instructor answers are single responses to a post formulated by students and instructors respectively. Each post can be followed by one or more follow-up discussions. Figure \ref{fig:piazza-post} shows a post on Piazza answered by a student and an instructor, along with a follow-up discussion.

\section{Background and Related Work}
    MOOC instructors report a decrease in work life balance, difficulty meeting expectations, and funding deficiencies compared to those teaching equivalent on-campus classes~\cite{Zheng2016}, and increased class sizes mean instructors are unable to provide the same level of student interaction, leading to higher student drop out rates~\cite{Kizilcec2013}. Existing research from Brinton \etal~\cite{Brinton2014} shows that the amount of threads and discussions produced quickly became intractable for both students and instructors to effectively navigate and search through. A correlation has been shown between whether students completed a MOOC and how active they were on the class's forums~\cite{Cohen2019}.
    
    Existing work in solving the problem of student-course interaction in growing online degree seeking programs include the use of intelligent agents, recommendation systems, and non-technology approaches. Jill Watson, a virtual teaching assistant that formulates answers to student questions, is one such intelligent agent~\cite{JillWatson2016}, using a hand-tuned semantic parsing algorithm to map student questions to concepts which can then be mapped to appropriate precompiled responses. Yang \etal~\cite{yang2014forum} use adaptive feature-based matrix factorization to present students with posts they would likely find interesting or participate in, helping to solve the scaling issue by further engaging students in question answering.
    
    In contrast, Joyner~\cite{Joyner2018} investigated non-technical approachs where teaching teams used different workflows to handle the scale of the class forum.
    
    PARQR is inspired by a class of recommender systems called Remembrance Agents, which retrieve information based on a user's current context. Remembrance Agents run as a continuous background process, rather than being actively invoked by a user~\cite{Rhodes2002Starner}. PARQR uses this method in online classrooms, detracting less from the user experience compared with switching modes to a keyword search. 

\section{PARQR}
    PARQR is most fundamentally an intelligent agent which aims to present relevant and actionable content to students and instructors based on the context the user is operating in. The design framework for determining when PARQR should provide content and what sort of content it should provide is simple: given the user is in a given \emph{context}, what information can we \emph{suggest} that helps focus the user on actionable content? When students arrive on Piazza's home page, PARQR assumes they want to browse previous posts to stay up to date with course discussions or gain insight on an assignment. Here PARQR suggests posts with large amounts of views and many follow up discussions --- indicators that other students have found these posts worthy of their attention. When students are asking questions, PARQR assumes they wish to find an answer to their question and presents related posts that may contain it. When instructors arrive on Piazza's home page, PARQR assumes they are there to answer questions for students, so it suggests questions that have not yet been answered but have received a large amount of student attention.
    
\begin{figure}[!htb]
\includegraphics[width=0.45\textwidth]{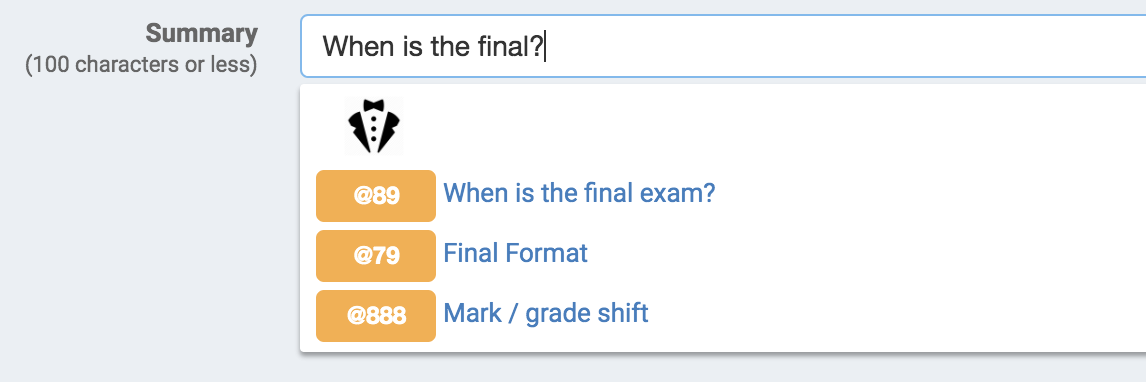}
\caption{\label{fig:piazza-question-suggestion} What a PARQR user sees when authoring a new question.}
\vspace{-5pt}
\end{figure}

%thad: figure out these #s
%fall2018 6601:
% 3337 total posts*
% 22984 total contributions**
% 2937 instructors' responses
%747 students' responses
%In CS6750 last term, we received 1533 individual threads and 15,367 individual contributions.
%In CS6460 last term, we received 785 individual threads and 15,597 individual contributions.
%In CS7637 last term, we received 1641 individual threads and 13,814 individual contributions.

\subsection{The Student's Experience}
When students arrive on the home page for a class in Piazza, PARQR suggests posts that class activity has indicated are attention worthy. In the same way that students benefit from questions asked in an in-person classroom, students here can benefit from questions asked in an online classroom, and we have the added ability to filter them based on how useful other students have indicated they are (i.e., the number of views they've received). We display the six questions from that class with the highest score defined by the equation:
    
    \begin{equation}
    \label{eqn:stud_rec_importance}
        I_n = \frac{v_n * f_n}{1 + e^{(a_n - \theta)}}
    \end{equation}
    where:
    \begin{equation_parameters}
     I_n     &  The importance of post $n$ \\
     v_n     &  Normalized number of views on post $n$ \\
     f_n     &  Normalized number of followups on post $n$ \\
     a_n     &  Age of post $n$ in days \\
    \end{equation_parameters}
    
We add a hard coded condition to never display posts that are older than three weeks or that do not have an instructor answer. The number of post views and followups are min-max normalized, and the parameter $\theta$ is used as an offset for the sigmoid method and set to seven days, which has the effect of prioritizing newer posts.

The second kind of suggestions that PARQR makes for students is to recommend existing, related posts as students are writing a new post (fig \ref{fig:piazza-question-suggestion}). The goal is to answer a student's question immediately (thereby preventing a duplicate post) or at least help better inform the question.

To make these recommendations, PARQR computes a distance from the vectorized version of the in-progress post (the post header, body, and selected tags) to each vectorized post in the database, where distance is the cosine similarity between the two posts. Then, PARQR recommends the five closest posts. This process is expanded upon in the ``architecture'' section.

\subsection{The Instructor's Experience}
When instructors visit the home page of a class in Piazza, PARQR shows them posts that we believe they would want to take action on: recently posted questions with a lot of views and followups but no instructor answer. When choosing these posts, PARQR first finds posts with no \emph{instructor response}. If there are more than six of these, it additionally filters by posts with no \emph{student response}. If there are still more than six, they are sorted in descending order by how many \emph{unresolved followups} they have, and the top six are reported.

\subsection{System Architecture}
In order to facilitate scaling up to a large amount of active concurrent users, we designed the backend system as a collection of micro services broken into separate Docker containers.

PARQR uses Faran's unofficial Piazza API~\cite{PiazzaAPI} to retrieve posts and populates our MongoDB with the post title, body, tags, answers, and any followups. The model training service runs sequentially after the parsing service to update models with new information. We use Sci-Kit Learn~\cite{scikit-learn} and the Natural Language Toolkit (NLTK)~\cite{BirdKleinLoper09} to create four models per class. These models are Term Frequency-Inverse Document Frequency (TF-IDF) encodings of 1) the concatenation of the post title, body, and tags 2) the answer from an instructor 3) the answer from a student and 4) the concatenation of all follow-up discussions on the question. Before these posts are encoded, we use the NLTK SnowballStemmer and WordNetLemmer to reduce the dimensionality of our data. The final models and embeddings of existing posts are cached in a Redis queue. This entire process occurs once per class every 15 minutes.

To render this information to the end-user, we built a browser extension which monitors the user activity and inserts recommendations into their page as discussed above. The browser extension is supported on both Google Chrome and Mozilla Firefox browsers. It communicates with a RESTful backend API service built with python using FLASK. When a student begins writing a question, the browser extension sends the contents of their question to the API. The API then transforms this text using the TF-IDF models from the corresponding class. Using cosine similarity as a distance metric, this vector is compared to all other vectorized posts from this class using the four models described above. The final score is computed as a weighted average of the scores from the four models- these weights were hand tuned with the highest weight given to the model trained on the content of the question. PARQR displays the five suggestions with the highest similarity. The browser extension allows us to monitor how PARQR is used, and registers events whenever a user clicks on the New Post button, one of our recommendations, or the Submit Post button.
   
\section{Experiments}
We determined that PARQR reduces the number of duplicate posts by comparing the Piazza forum activity for a particular course taught with and without the use of PARQR. For our analysis we chose the Introduction to AI course during the Spring 2017 and Spring 2019 semesters and focus on the time period of Assignment 2, in which students implement multiple classical AI search algorithms. Though PARQR is used by multiple classes, we selected the Introduction to AI class for our experiments because it has a large number of students who are very active on Piazza. Assignment 2 was chosen because is identical in subsequent semesters and is early enough in the semester that most students who would drop were still in the class. Additionally, this class consists entirely of online degree seeking students. Spring 2017 was prior to any development of PARQR and thus had no PARQR users. In Spring 2019 PARQR was used by 98\% of Piazza users in the class.

\subsection{Duplicate Post Inter-Rater Reliability}
Here we define two or more questions as duplicates if a student writing one of them would stop upon being shown answers to any of the others (i.e., the posts contain sufficient information to answer one another). This judgment is subjective, and the size of the dataset is large enough that no one researcher could be responsible for labeling it. We isolated two sets of posts: Assignment 1 from Spring 2017 for calculating interrater reliability and Assignment 2 from both semesters for analyzing PARQR's impact.

Three researchers hand clustered assignment 1 from the Spring 2017 course. The researchers were given a spreadsheet of Piazza posts from Assignment 1, including the subject, body, and answers to the posts. Each researcher clustered the posts into groups of duplicates. The researchers then negotiated and merged their clusterings to create a single ``gold standard'' clustering of the posts.

Next, three other researchers were given 200 pairs of questions pulled from this gold standard in which the ratio of pairs which were duplicates (17\%) matched that of the gold standard. They were asked to label each of these as either a duplicate or not. All three raters, acting independently, agreed on whether two questions are duplicates 90\% of the time. The average agreement of the three combinations of two inter-rater comparisons was 93\%. This is significantly above the baseline agreement of 83\%, which could have been attained by always predicting the majority class (not duplicate) Given these results, we are confident in having multiple researchers cluster the larger datasets, as done in the next section.

\subsection{Analysis of Reduction of Duplicate Posts}

To determine the effect of PARQR on duplicate posts, we shuffled posts from the Spring 2017 and Spring 2019 assignment 2 into one dataset, keeping the labelers blind to which semester each post came from, and storing a mapping to allow reversing this process. Eight researchers then collaboratively and asynchronously clustered this data into clusters of duplicate posts.  Note that since the semesters are combined, the defined clusters were common across both. The data was then split back into their original semesters. The results of this analysis are summarized in Table \ref{tab:multi_year}.

%   Here our aim is not to show the accuracy of the raters in determining if a question pair is a duplicate, but rather that they agree on their choices by more than chance.
    \begin{table}
      \centering
      \begin{tabular}{p{4cm} r r}
        % \toprule
        {\textit{Statistic}} 
        & {\textit{Spring 2017}}  
        & {\textit{Spring 2019}} \\
        \midrule
        Enrolled students & 390 & 590 \\
        Students active on Piazza & 385 & 590 \\
        PARQR Users (Percentage) & 0 (0\%) & 578 (98.0\%) \\
        Number of posts & 195 & 168\\
        Number of duplicate posts & 50 (25.6\%) & 30 (17.8\%) \\
        Posts per active student & 0.506 & 0.291 \\
        %Unique posts per active student & XX & ZZ \\
        %Duplicate posts per active student & 0.129 & 0.05 \\
%        Percentage of duplicate posts & 25.6\% & 17.8\% \\
        % \bottomrule
      \end{tabular}
      \caption{Piazza metrics during Assignment 2.}~\label{tab:multi_year}
      \vspace{-15pt}
    \end{table}
%thad: change statistics to be based only on active Piazza users
%thad: potentially add \# of contributions and \# of contributions on duplicate threads 

The metrics gathered in Table \ref{tab:multi_year} show a drop in the number of duplicate posts from Spring 2017 to Spring 2019. We performed a one-sided Z-test on the difference in these proportions and found the drop in duplicate posts statistically significant ($p=0.0392$). Note also that the number of posts per active student declined.

\subsection{Evaluating the Model}
To evaluate the performance of our model, we performed a walk-forward validation experiment. We used the labelled gold standard dataset from Assignment 1 of the Fall 2017 AI course containing 179 posts, of which 34 posts had at least one prior duplicate post. We ordered the posts chronologically. Then, for each post $p_i$, we trained our model ensemble on posts $p_0$ to $p_{i-1}$ and queried the model for relevant posts to post $p_i$. The model was able to retrieve at least one relevant post, if it existed, 73.5\% of the time.

\subsection{Instructor Interviews}
Five teaching assistants (TA's) from the Spring 2019 Introduction to AI teaching staff were interviewed, all of whom had previously taken the class. Four of the five TA's used PARQR as instructors. TA's reported that the posts that were suggested to them on their home page as ``high attention posts'' were a useful way to keep on top of new posts and as a way to gauge which questions or topics were likely to be asked about during office hours (which each TA holds once a week). One TA noted that because the instructor suggested posts display the number of follow ups, they can tell when there is an active discussion going on. When asked about whether or not PARQR reduced the number of duplicate posts in a class, instructors were unsure, suggesting that instructors will be unable to tell if PARQR is having an overall impact on the student experience in the class.

\section{Future Work}
We are continuing to interview more instructors and students in order to gain a better understanding of PARQR's impact and users' experiences with it. We are working to capture more information from student activity so we can understand their behavior on Piazza and how PARQR changes this behavior, especially over time. We continue to look for new ways to use PARQR to augment the online classroom experience with tools and supplemental information that help students and instructors alike.

% Balancing columns in a ref list is a bit of a pain because you
% either use a hack like flushend or balance, or manually insert
% a column break.  http://www.tex.ac.uk/cgi-bin/texfaq2html?label=balance
% multicols doesn't work because we're already in two-column mode,
% and flushend isn't awesome, so I choose balance.  See this
% for more info: http://cs.brown.edu/system/software/latex/doc/balance.pdf
%
% Note that in a perfect world balance wants to be in the first
% column of the last page.
%
% If balance doesn't work for you, you can remove that and
% hard-code a column break into the bbl file right before you
% submit:
%
% http://stackoverflow.com/questions/2149854/how-to-manually-equalize-columns-
% in-an-ieee-paper-if-using-bibtex
%
% Or, just remove \balance and give up on balancing the last page.
%

% BALANCE COLUMNS
\balance{}

% REFERENCES FORMAT
% References must be the same font size as other body text.
\bibliographystyle{SIGCHI-Reference-Format}
\bibliography{references}

\end{document}